\documentclass[aps,twocolumn,prl,showpacs,showkeys,array]{revtex4}
\usepackage[dvips]{graphicx}
\usepackage{psfrag}

\begin{document}

\title{Resonant scattering of solitons}
\author{A. E. Miroshnichenko, S. Flach}
\affiliation{Max-Planck-Institut f\"ur Physik komplexer Systeme, N\"othnitzer
Strasse 38, D-01187 Dresden, Germany}
\author{B. Malomed}
\affiliation{Department of Interdisciplinary Studies,
Faculty of Engineering , Tel Aviv University
Ramat Aviv 69978, Israel}
\date{\today}

\begin{abstract}
We study the scattering of solitons in the nonlinear Schr\"odinger equation
on local inhomogeneities which may give rise to resonant transmission and
reflection. In both cases, we derive resonance conditions for the soliton's
velocity. The analytical predictions are tested numerically in regimes
characterized by various time scales. Special attention is paid to intermode
interactions and their effect on coherence, decoherence and dephasing of
plane-wave modes which build up the soliton.
\end{abstract}

\maketitle
{\bf The scattering of solitons by various scattering centers
in the nonlinear Schr\"odinger equation leads to resonant
transmission and reflection if the soliton velocity matches
certain resonance conditions. By assuming that the soliton
is composed of a weighted superposition of modes (i.e. a
wave packet) different scattering regimes are observed depending
on the ratio of the duration of the scattering event and
the characteristic mode-mode interaction time due to nonlinearity.
Resonant transmission does not suffer from mode dephasing,
while resonant reflection (Fano resonance) does. Consequently
transmission resonances are observed independent of the
scattering regime, while the opposite holds
for Fano reflection resonances.}

\section{Introduction}

Transmission of linear waves through inhomogeneous media is a topic of wide
interest \cite{psbwzzgp89,Klyatskin,LifGredPast,Zwerger,dhgpt99}. In the
one-dimensional case, two basic resonant features of wave scattering are
known. One of them corresponds to the wave travelling above a barrier or a
potential well. Proximity of the wave's frequency to that of a standing wave
captured inside the scattering potential leads to a resonance, i.e., the
transmission may be strongly enhanced in a vicinity of the resonance \cite
{LLIII}. In fact, the transmission at resonance is perfect, which is used in
Fabry-Perrot interferometers \cite{cohen}. The connection of the resonant
scattering to the Levinson's theorem and the existence of bound states in
potential wells is discussed, e.g., in Ref. \cite{LLIII}.

Resonant reflection is possible too, as a consequence of the Fano resonance
\cite{fano}, and it occurs whenever a wave may choose between two scattering
paths, which finally merge into one exit. Destructive interference may then
lead to total reflection. An adequate explanation goes through the coupling
of a local Fano state to a continuum \cite{gdm93,gl93,ns94}. The wave may
then again either pass directly through the continuum or by visiting the
Fano state. The resonance condition is provided directly by the matching of
the wave's frequency to the eigenfrequency of the Fano state. More than one
wave paths are typically also generated by time-dependent scattering
potentials.

Fano resonances imply the coherence of the wave phases throughout the
scattering process, hence any significant dephasing effects will suppress
the resonance. At the same time, the resonant transmission mechanism does
not rely on phase coherence, therefore dephasing does not destroy it.


Similar situations can also be observed in nonlinear systems which possess
continuous or discrete translational invariance. For instance, nonlinear
lattice models support time-periodic spatially localized states in the form
of discrete breathers \cite{breather}. If small-amplitude plane waves are
sent towards such a breather, it acts as a time-periodic scattering
potential. The temporal periodicity of the potential leads to excitation of
many new scattering channels, shifted by multiples of the breather's
fundamental frequency relative to the frequency of the incoming wave \cite
{scattering2}. Even if all the additional channels are closed ones, i.e.,
they do not match the plane-wave's frequency, they may generate localized
Fano states which resonate with the open channel and thus lead to
Fano-resonant backscattering \cite{sfaemvfmvf02}. In addition, resonant
transmission can also be observed in this case \cite{skcbrsm97}.

In this paper we study the transmission properties of small-amplitude
solitons (rather than plane linear waves) in the discrete nonlinear
Schr\"{o}dinger (DNLS) equation with two types of scattering centers. The
first one is a Fano-defect center, which is an extra level coupled to the
DNLS equation at one site of the underlying lattice. This Fano center may
actually be a strongly localized discrete breather of the DNLS model \cite
{sfaemvfmvf02}. The second type is a two-site impurity which gives rise to
two bound states. Our goal is to predict resonant backscattering and
resonant transmission of the small-amplitude soliton impinging onto these
defects. To this end, we consider the soliton as a superposition of plane
waves or modes, while the nonlinearity leads to mode-mode interactions,
which may or may not cause dephasing effects in the course of the scattering
process. We will find values of the velocity of the incident soliton at
which the resonant transmission and reflection are possible.

The predicted effects can be observed in any physical system which is
modelled by the DNLS equation, the most straightforward ones being arrays of
nonlinear optical waveguides. These may be realized as set of parallel cores
fabricated on a common substrate \cite{Yaron}, or a virtual array induced in
a photorefractive material illuminated by a set of parallel laser beams \cite
{Moti}.

\section{Resonances in the scattering of linear waves}

Before considering the scattering of solitons, we will set the stage, using
two simple models which make it possible to observe resonant transmission
and reflection in linear wave scattering. In both cases, we will use the
linear discrete Schr\"{o}dinger equation as the unperturbed system. Later
on, we will add nonlinearities to make propagation of solitons possible.

\subsection{Resonant transmission}

Resonant transmission due to bound states is possible if the scattering
potential supports at least one bound state. To be flexible, we take a
lattice model which allows for two or one bound states, depending on its
parameters:
\begin{equation}
i\dot{\phi}_{n}=C(\phi _{n-1}+\phi _{n+1})+\delta _{n,0}\epsilon _{0}\phi
_{n}+\delta _{n,1}\epsilon _{1}\phi _{n}\;.  \label{restran}
\end{equation}
Here $\phi _{n}$ is the complex scalar dynamical variable at the $n$-th
site, a real constant $C$ controls the strength of the inter-site coupling,
and two diagonal defects are set at sites $n=0$ and $n=1$ with strengths
$\epsilon _{0}$ and $\epsilon _{1}$, respectively. In the absence of the
defects, Eq. (\ref{restran}) has exact plane-wave solutions, $\phi _{n}=
\mathrm{\exp }\left[ i(\omega _{q}t-qn)\right] $, whose frequency satisfies
the dispersion relation
\begin{equation}
\omega _{q}=-2C\cos q\,.  \label{spectrum}
\end{equation}

\begin{figure}[htb]
\vspace{20pt} \centering
\includegraphics[width=0.450\textwidth]{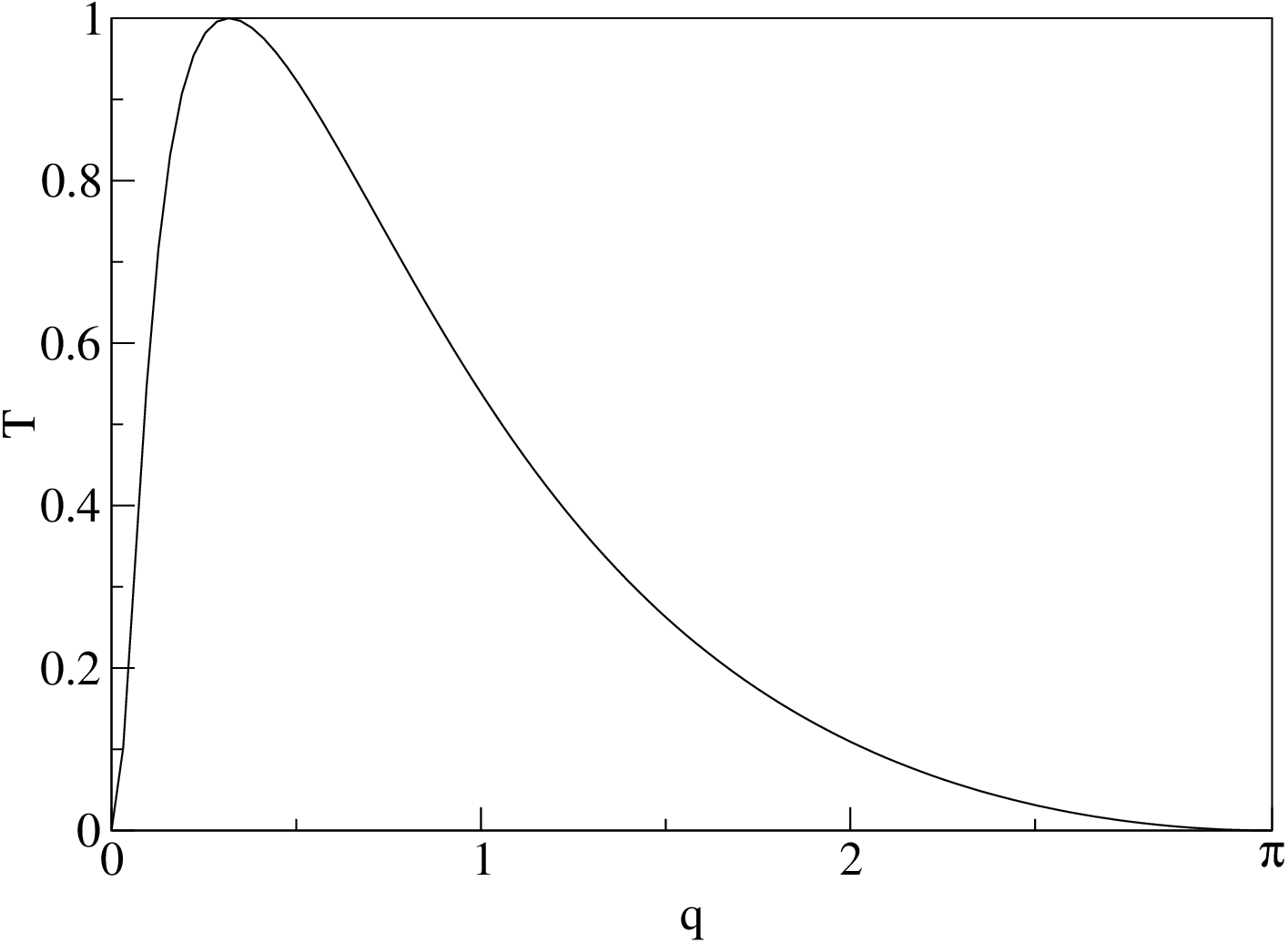}\newline
\caption{The transmission coefficient $T$ versus $q$ for the case of the
resonant transmission, as per Eq. (\ref{tr2}) with $C=10$ and
$\epsilon=19$.}
\label{fig6}
\end{figure}

To find the transmission coefficient $T$ in the presence of the two diagonal
defects, we impose proper boundary conditions and obtain, after some algebra
(see, e.g., Ref. \cite{BambiHu}), for the case of $\epsilon _{0}=\epsilon
_{1}\equiv \epsilon $,
\begin{equation}
T=\frac{4C^{4}\sin ^{2}q}{\epsilon ^{2}(\epsilon -2C\cos q)^{2}+4C^{4}\sin
^{2}q}\;.  \label{tr2}
\end{equation}
Perfect transmission ($T=1$) is possible in the case $\epsilon \leq 2C$
under the resonance condition (at the special value of $q$),
\begin{equation}
q^{\prime }=\cos ^{-1}(\epsilon /2C)\;.  \label{qrestran}
\end{equation}
For small values of the impurity strength, $\epsilon \ll 2C$, the perfect
transmission takes place around $q=\pi /2$. With $\epsilon $ approaching
$2C$, the value of $q^{\prime }$ moves towards zero (see Fig.\ref{fig6}).
At $\epsilon =2C$, the perfect transmission occurs exactly at the edge point of
the spectrum, $q^{\prime }=0$, and disappears for larger values of $\epsilon
$. Therefore, it is possible to control the location of the resonance by
varying the impurity strength $\epsilon $.

%

\subsection{Resonant reflection}

In order to produce a Fano resonance, we consider the Fano-Anderson model,
which amounts to adding an additional local Fano degree of freedom $\varphi$,
with the eigenfrequency (energy) $E$, to the linear discrete
Schr\"{o}dinger equation. The dynamical variable $\varphi $ is coupled to
the lattice field $\phi _{n}$ at the site $n=0$:
\begin{eqnarray}
i\dot{\phi}_{n} &=&C(\phi _{n-1}+\phi _{n+1})+\epsilon \varphi \delta
_{n0}\;,  \nonumber \\
i\dot{\varphi} &=&-E\varphi +\epsilon \phi _{0}\;.  \label{eq2}
\end{eqnarray}
When $\epsilon =0$, the system decouples into the free wave with the
spectrum $\omega _{q}=-2C\cos q$ and an additional localized level with the
energy $E$. The value of $E$ is chosen so that it lies inside the continuous
spectrum, i.e., $|E|<2C$.

If $\epsilon \neq 0$, the Fano defect interacts with the continuous spectrum
locally. To solve the linear system (\ref{eq2}), we substitute
\[
\phi _{n}=A_{n}\exp \left( i\omega _{q}t\right) \,,\,\varphi =B\exp \left(
i\omega _{q}t\right) \,,
\]
and obtain
\begin{eqnarray}
-\omega _{q}A_{n} &=&C(A_{n-1}+A_{n+1})+\epsilon B\delta _{n0}\;,  \nonumber
\\
-\omega _{q}B &=&-EB+\epsilon A_{0}\;.  \label{AB}
\end{eqnarray}
We use the second equation in (\ref{AB}) to eliminate $B$ in favor of
$A_{0}$, then the first equation yields
\begin{equation}
-\omega _{q}A_{n}=C(A_{n-1}+A_{n+1})-\epsilon ^{2}\left( \omega
_{q}-E\right) ^{-1}A_{0}\delta _{n0}\;.  \label{eq5}
\end{equation}
Equation (\ref{eq5}) amounts to the presence of a \textit{resonant}
scattering potential, which depends on the frequency of the incident wave.
Moreover, if the energy of the additional level is located inside the
continuous spectrum, $|E|<2C$, as it was assumed above, there is a value
$q_{F}=\cos ^{-1}\left( -E/2C\right) $ at which the denominator of the last
term in Eq. (\ref{eq5}) vanishes, according to Eq. (\ref{spectrum}). At this
value of $q$, complete reflection, $T=0$, takes place, as the scattering
potential becomes infinitely large.

\begin{figure}[tbh]
\vspace{20pt} \centering
\includegraphics[width=0.45\textwidth, height=4.5cm]{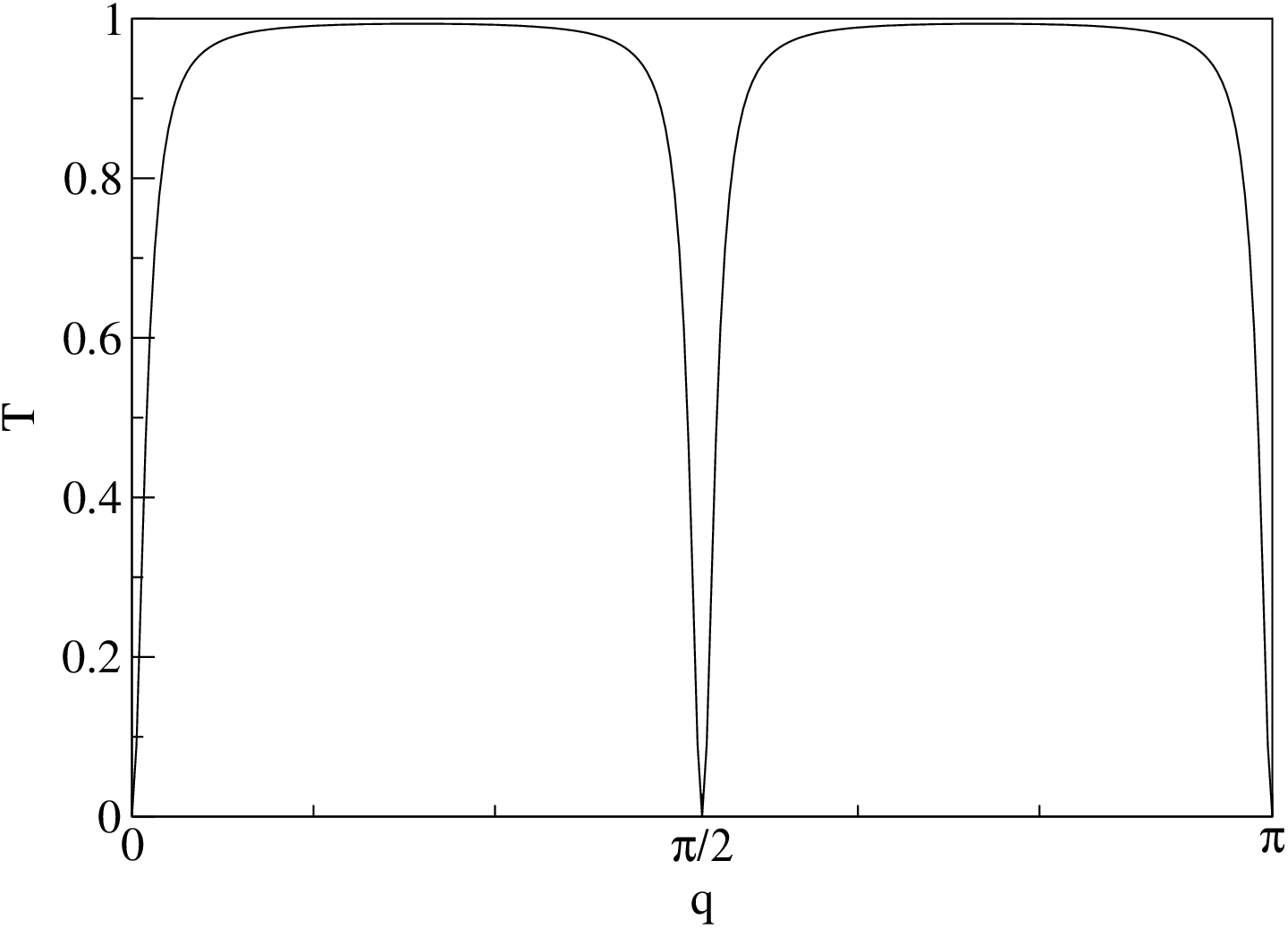}\newline
\caption{The transmission coefficient as given by Eq. (\ref{eq9}), versus
the wavenumber $q$ of the inident wave. The parameters are $E=0$, $C=10$
(i.e., $q_{F}=\protect\pi /2$), and $\protect\epsilon =4$.}
\label{fig1}
\end{figure}

After some algebra (see, e.g., Ref. \cite{BambiHu}), the transmission
coefficient $T$ for the linear model based on Eqs. (\ref{eq2}) can be
written as
\begin{equation}
T=\left[ 1+\frac{1}{4\sin ^{2}q}\frac{\epsilon ^{4}}{C^{2}(E-\omega _{q})^{2}
}\right] ^{-1}\;.  \label{eq9}
\end{equation}
A typical dependence of $T$ versus the incident wavenumber $q$ is shown in
Fig. \ref{fig1} for $E=0$.
%
In the case of $E=0$, the transmission coefficient $T(q)$ vanishes at
$q_{F}=\pi /2$. The additional vanishing of $T$ at the band edges $q=0$ and
$q=\pi $ is related to the fact that the group velocity of plane waves is
equal to zero at these points, and will not be discussed below.

If $\epsilon ^{2}/\left( 2C^{2}\right) \ll \sin q$, then $T\approx 1$ at all
values of $q$ except for close to $q=q_{F}$, as $T(q_{F})=0$. The width of
the resonance in the $q$-space is defined by the distance between points at
which $T=1/2$, or, as it follows from Eq. (\ref{eq9}),
\begin{equation}
\frac{\epsilon ^{4}}{4C^{2}(E-\omega _{q})^{2}\sin ^{2}q}=1\;.  \label{eq10}
\end{equation}
By expanding $\omega _{q}$ around $q_{F}$, $\omega _{q}\approx \omega
_{q_{F}}+\omega _{q}^{\prime }(q_{F})\delta q$, and substituting this into
Eq. (\ref{eq10}), we find
\begin{equation}
\delta q=\frac{\epsilon ^{2}}{4C^{2}\sin ^{2}q_{F}}\;.  \label{eq13}
\end{equation}
The width of the resonance is twice the expression (\ref{eq13}),
\begin{equation}
\Delta q=\frac{\epsilon ^{2}}{2C^{2}\sin ^{2}q_{F}}\;.  \label{eq14}
\end{equation}

The above considerations are valid also for the model based on the
continuous linear Schr\"{o}dinger equation augmented with a corresponding
$\delta $-like scattering term. Quite a different situation takes place if
the local scattering term, added to the linear Schr\"{o}dinger equation, is
nonlinear -- in that case, the solution to the scattering problem is not
unique, and may be subject to local modulational instability
\cite{MarkAzbel}.

\section{Soliton scattering}

In order to consider a possibility of resonant transmission or reflection in
the scattering of a soliton on a local defect, we first add nonlinear terms
to the linear Schr\"{o}dinger equation [see the first equation in the system
(\ref{eq2})], thus arriving at the DNLS equation,
\begin{equation}
i\dot{\phi}_{n}=C(\phi _{n-1}+\phi _{n+1})+\lambda |\phi _{n}|^{2}\phi
_{n}\;.  \label{dnls}
\end{equation}
Small-amplitude (hence, broad) moving solitons in Eq. (\ref{dnls}) are
approximated well by the corresponding solution to the continuous NLS
equation,
\begin{eqnarray}
\phi _{n}(t)&\approx& \frac{\eta }{2}\sqrt{\frac{\lambda }{2C}}\mathrm{\exp }
\left[ i(\frac{V}{2C}n-(\omega _{s}+2)Ct)\right]\times\nonumber\\
&&\mathrm{sech}\left[ \frac{
\eta \lambda }{4C}(n-Vt)\right] ,  \label{eq16}
\end{eqnarray}
where $V$ is the velocity of the soliton, $\omega _{s}=\left( 4C^{2}\right)
^{-1}(V^{2}-\eta ^{2}\lambda ^{2}/4)$ is its intrinsic frequency, and $\eta$
is the amplitude \cite{yskbam89}.

The velocity $V$ determines a central wavenumber of the soliton's Fourier
transform,
\begin{equation}
q_{c}=V/\left( 2C\right) \,.  \label{q_c}
\end{equation}
The soliton may be considered as a superposition of linear plane waves with
wavenumbers taking values in an interval around $q_{c}$, the width of this
interval being
\begin{equation}
\Delta q_{c}=\ln \left( 2+\sqrt{3}\right) \frac{\eta \lambda }{\pi C}\;.
\label{eq17}
\end{equation}
This spectral width is to be compared with the width of the corresponding
resonances [e.g., with the expression (\ref{eq14})].

There are two characteristic time scales, which describe the scattering of
the soliton (\ref{eq16}) by the resonant defect. One of them is the time of
the soliton-defect interaction, which is estimated as the soliton's width
$1/\Delta q_{c}$ in the coordinate space, divided by its velocity:
\begin{equation}
\tau _{\mathrm{int}}=\left( \Delta q_{c}V\right) ^{-1}\;.  \label{eq18}
\end{equation}
The interaction between waves composing the free soliton defines the second
time scale. It can be estimated as the time of dispersion of the wave packet
(\ref{eq16}) in the linearized equation (\ref{dnls}), with $\lambda =0$,
which yields, similar to Eq. (\ref{eq18}), a result
\begin{equation}
\tau _{\mathrm{disp}}=\left( \Delta q_{c}\Delta v_{g}\right) ^{-1}\;,
\label{eq19}
\end{equation}
where $v_{g}(q)=2C\sin q$ is a group velocity of the waves, and
\begin{equation}
\Delta v_{g}=v_{g}(q_{c}+\Delta q_{c}/2)-v_{g}(q_{s}-\Delta q_{c}/2)=4\sin
\left( \Delta q_{c}/2\right) \cos q_{c}  \label{Delta}
\end{equation}
is the relative velocity between faster and slower waves composing the
soliton.

The interaction
between the plane-waves constituents does not play a significant role during
the scattering of the soliton if
\begin{equation}
\tau _{\mathrm{int}}\ll \tau _{\mathrm{disp}}\,,
\label{timescales}
\end{equation}
hence, under this condition, the soliton may be considered as a set of
noninteracting plane waves while it suffers scattering on the defect.
The transmission of each wave component is then
determined by the corresponding coefficient for the linear model, see the
previous section. If, in this regime, $\Delta q_{c}$ is sufficiently
small in comparison with the width of the transmission or reflection
resonance for the plane waves, then all the waves composing the soliton will
be resonantly transmitted or reflected, provided that velocity $V$ matches
the resonance condition.

When $\tau _{\mathrm{int}}\gg \tau _{\mathrm{disp}}$, the wave-wave
interactions become important during the scattering process. Since these
interactions may lead to dephasing of individual waves, we expect that the
resonant transmission will not be affected, while the resonant Fano
reflection will be suppressed, as it relies on keeping the wave phase
coherence in the course of the scattering process.

\subsection{Resonant reflection of solitons}

A nonlinear model which may give rise to the resonant reflection of the
soliton is based on a straightforward generalization of Eqs. (\ref{eq2}),
\begin{eqnarray}
i\dot{\phi}_{n} &=&C(\phi _{n-1}+\phi _{n+1})+\lambda |\phi _{n}|^{2}\phi
_{n}+\epsilon \varphi \delta _{n0}\;,  \nonumber \\
i\dot{\varphi} &=&-E\varphi +\epsilon \phi _{0}\;.  \label{eq15}
\end{eqnarray}
We performed numerical simulations of Eqs. (\ref{eq15}) using the
fourth-order Runge-Kutta method. The total number of sites was large,
$N=2000 $. Time was measured in dimensionless units.

\begin{figure}[tbh]
\vspace{20pt} \centering \psfrag{f}{$10^2|\phi_n|^2$}
\includegraphics[width=0.450 \textwidth]{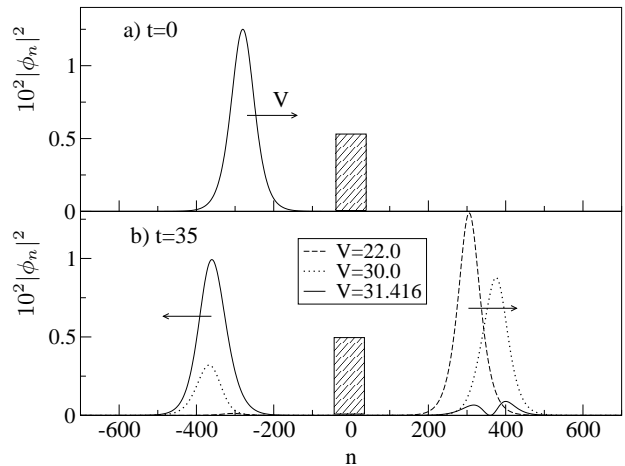}\newline
\caption{Profiles of the wave field in the initial state (a) and
after the interaction (b). The Fano defect (marked by the shaded
bar) is placed at $n=0 $, and the direction of motion of the
pulses is indicated by arrows. The parameters are: $E=0$ (i.e.,
$q_{F}=\frac{\protect\pi }{2}$), $C=10.0$, and $\protect\epsilon
=4.0$, $\protect\eta =\protect\lambda =1.0$. After the
interaction, two soliton-like pulses are observed, which move in
the opposite directions with equal absolute values of their
velocities.} \label{fig2}
\end{figure}

The soliton is launched from the left with a positive velocity $V$. After
the interaction with the Fano defect, the soliton is typically found to be
split into two soliton-like fragments which move in opposite directions, see
Fig.\ref{fig2}.
%
The transmission coefficient can be found from numerical data, using the
conservation of the norm $\sum\limits_{n}|\phi _{n}|^{2}$ in the DNLS
equation (another conserved quantity is the Hamiltonian). The transmission
coefficient is then defined as the ratio of the norms of the transmitted
wave packet and initial soliton:
\begin{equation}
T=\frac{\sum\limits_{n>0}|\phi _{n}(t^{\ast })|^{2}}{\sum\limits_{n}|\phi
_{n}(0)|^{2}}  \label{T}
\end{equation}
where we choose $t^{\ast }\gg \tau _{\mathrm{int}}$.

In the case shown in Fig. \ref{fig2}, we chose $q_{F}=\pi /2$. In this case,
the dispersion time (\ref{eq19}) diverges, as it follows from Eq. (\ref
{Delta}), and in the first approximation the soliton does not disperse at
all, even in the linear system. Further, the condition $\Delta q_{c}<\Delta
q $, which implies that the soliton's spectral size is smaller than the
width of the resonance [see Eq. (\ref{eq14})], hence all the spectral
components are expected to be resonantly reflected, is
\begin{equation}
\eta \lambda <\frac{\epsilon ^{2}}{C\sin ^{2}q_{F}}  \label{eq21}
\end{equation}
To satisfy this condition, other parameters were taken as $\eta =\lambda =1$,
$C=10$, and $\epsilon =4$.

\begin{figure}[tbh]
\vspace{20pt} \centering
\includegraphics[width=0.450\textwidth]{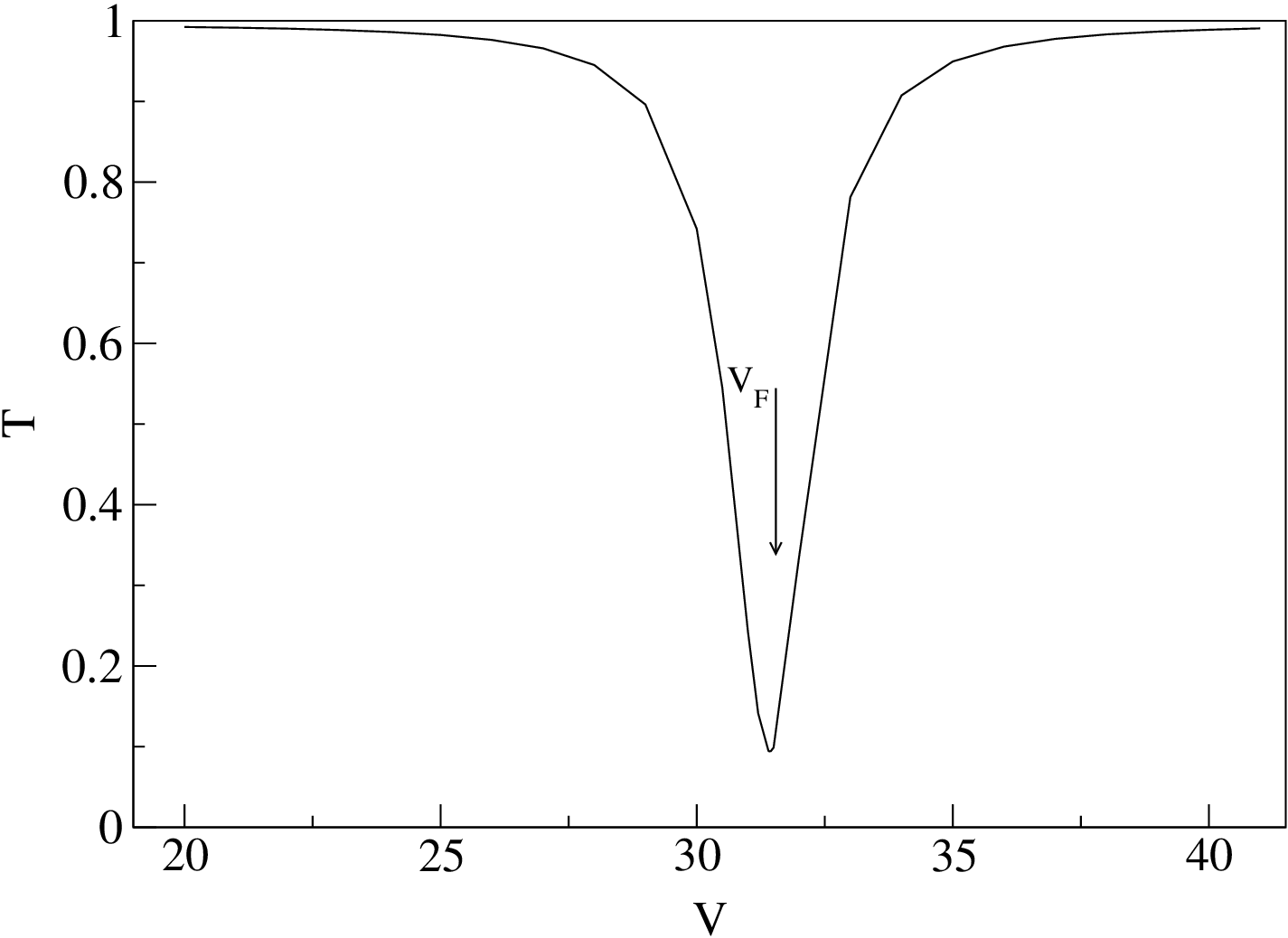}\newline
\caption{The transmission coefficient for the scattering of the soliton on
the Fano defect, defined as in Eq. (\ref{T}) and found from direct
simulations, versus the soliton's velocity $V$. Parameters are the same as
in Fig. \ref{fig2}. The minimum is located at $V_{F}=2Cq_{F}$.}
\label{fig3}
\end{figure}

The numerically computed transmission coefficient is shown, as a function of
the soliton's velocity, in Fig.\ref{fig3}.

%
There is a minimum in the dependence of the transmission on soliton velocity
exactly at $V_{F}=2Cq_{F}$, as it is predicted by Eq. (\ref{q_c}) and the
concept proposed above, according to which the soliton may be treated as a
spectrally narrow packet of quasi-linear waves, hence the resonant
backscattering should take place when the central wavenumber $q_{c}$
coincides with the linearly predicted Fano-resonance wavenumber, $q_{F}$.
The minimum does not reach zero, which is explained by the finite spectral width
$\Delta q_{c}$ of the soliton, i.e., wave components which build up
the soliton do not fully backscatter. It should be stressed that the
close proximity of the numerically found minimum to the anticipated point
predicates upon the conditions (\ref{eq21}) and (\ref{timescales}), but it
is not related to the fact that the particular example displayed in Fig.
\ref{fig3} has $q_F=\pi/2$. In fact, numerical results are very similar
at other values of $q_F$.

To study the case when the wave-wave interactions are important during the
time of the interaction between the soliton and the defect, i.e., $\tau _{
\mathrm{int}}\,_{\sim }^{>}\,\tau _{\mathrm{disp}}$, the dispersion time was
changed by shifting the soliton's central wavenumber $q_{c}$ to the
spectrum's edge, keeping all other parameters fixed. The Fano resonance is
observed in the simulations until $\tau _{\mathrm{int}}\approx \tau _{
\mathrm{disp}}$. With the further decrease of $\tau _{\mathrm{int}}$, the
Fano minimum in the $T(V)$ curve becomes less pronounced, and disappears at
some critical value of $q_{c}$. For instance, we have found numerically that
for $C=\eta =\lambda =\epsilon =1$, when the condition of\ the smallness of
$\tau _{\mathrm{disp}}$ is fulfilled, the Fano resonance disappears at
$q_{c}\approx 1.0$.
%
\begin{figure}[tbh]
\vspace{20pt} \centering
\includegraphics[width=0.450\textwidth]{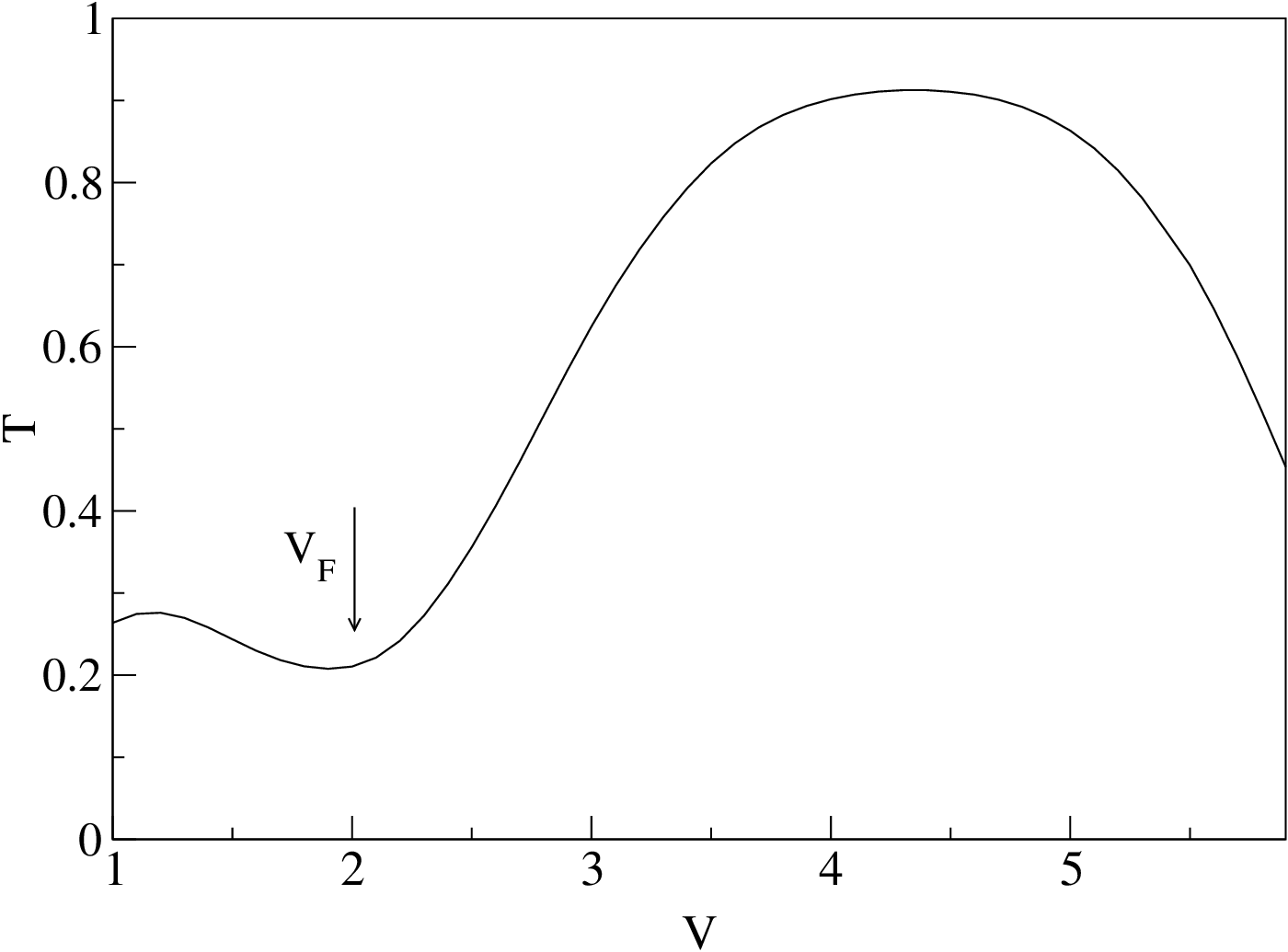}\newline
\caption{The dependence of the soliton's transmission coefficient in the
case of the scattering on the Fano defect on its velocity $V$. The
parameters are $C=\protect\eta =\protect\lambda =\protect\epsilon =1.0$, and
$q=1$. The minimum at $V_{F}=2Cq_{F}$ is quite shallow in this case.}
\label{fig4}
\end{figure}

Still we observe a smooth $T(V)$ curve (Fig. \ref{fig4}). This result
implies that the plane-wave modes which make up the soliton loose their
resonant backscattering features. This is not unexpected, as the interaction
of many modes between themselves can be interpreted as a multichannel
scattering problem (as opposed to many uncoupled single-channel scattering
processes in the absence of the mode-mode interaction). Multichannel
scattering problems are expected to have less pronounced Fano resonance
features, since the Fano resonance is inherently based on keeping phase
coherence in order to support the destructive interference. As the phase
coherence gets suppressed by the mode-mode interactions, the necessary
interference is detuned.

\subsection{Resonant transmission of solitons}

The passage of a soliton through a single-site impurity has been already
studied in many works \cite{oneimp}. Here we consider the nonlinear
extension of the two-site-impurity system (\ref{restran}),
\begin{equation}
i\dot{\phi}_{n}=C(\phi _{n-1}+\phi _{n+1})+\lambda |\phi _{n}|^{2}\phi
_{n}+(\delta _{n,0}+\delta _{n,1})\epsilon \phi _{n}\;.  \label{nrestran}
\end{equation}
%
\begin{figure}[tbh]
\vspace{20pt} \centering
\includegraphics[width=0.450\textwidth]{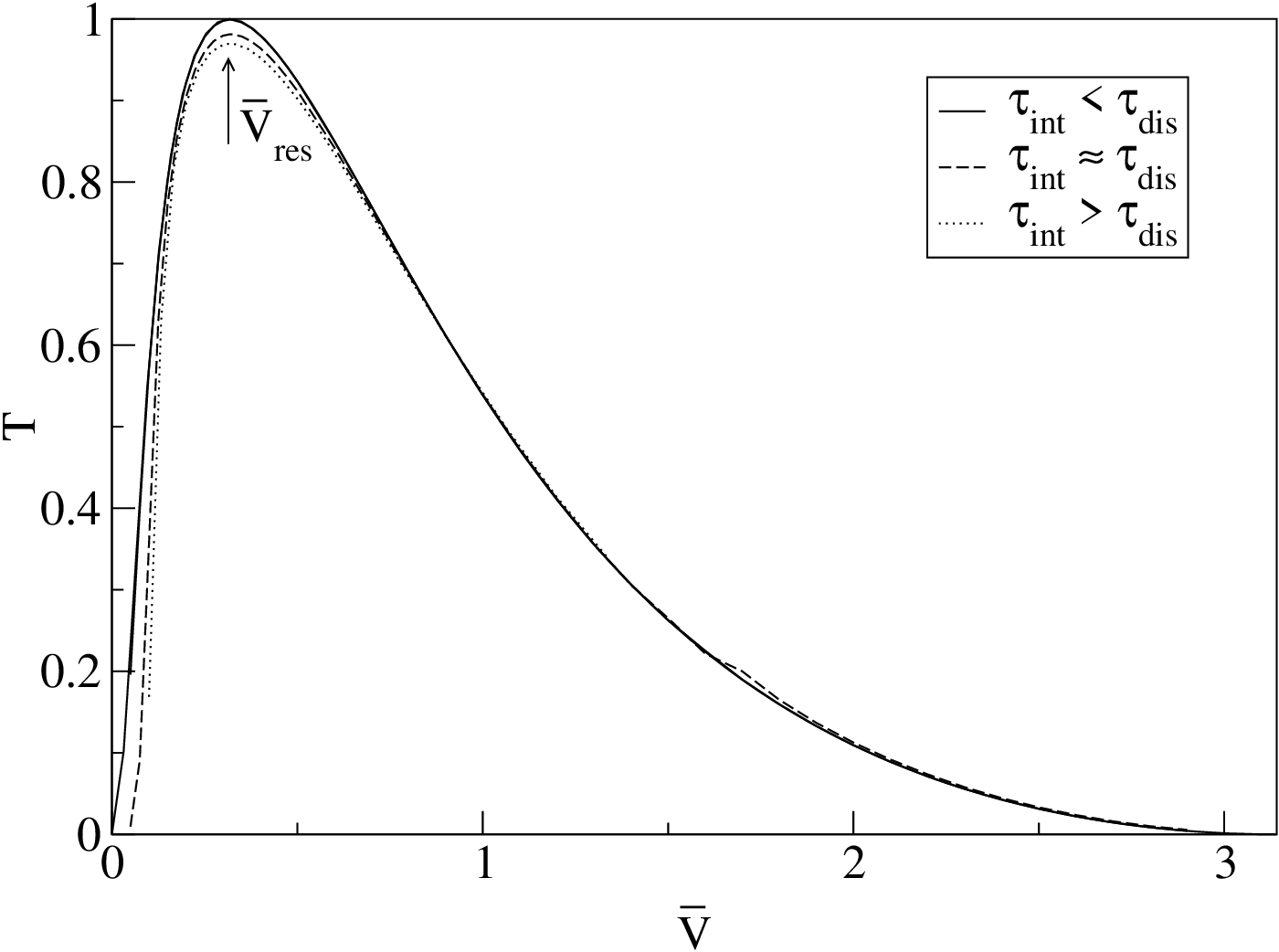}\newline
\caption{A typical dependence of the transmission coefficient $T$ for the
soliton in the model (\ref{nrestran}) on $V$ for $\protect\eta =
\lambda =1.$ and (a) $\protect\epsilon =19$ and $C=10$ (solid line); (b)
$\protect\epsilon =3.8$ and $C=2$ (dashed line); (c) $\protect\epsilon =2.85$
and $C=1.5$ (dotted line) In this figure the normalized velocity $\bar{V}=
\frac{V}{2C}$ is used. In all the cases, the position of the resonance is
fixed at $\bar{V}_{res}\approx 0.318$.}
\label{fig5}
\end{figure}
The transmission coefficient for the soliton passing this defect was
computed numerically for various values of parameters, and found to be in a
very good agreement with Eq. (\ref{tr2}), see Fig.\ref{fig5}. Remarkably,
neither the position of the resonance nor the value of the transmission
coefficient at the resonance is conspicuously affected when the wave-wave
interactions become important. This clearly indicates that dephasing does
not seriously harm the resonant-transmission mechanism.

\section{Conclusions}

We have shown that solitons may be resonantly transmitted or backscattered
depending on their velocity and the type of the scatterer. We explained
these effects by considering the soliton as a superposition of plane-wave
modes in the limit when the mode-mode interaction is not affecting the
scattering process. By tuning parameters into a regime where the mode-mode
interaction becomes essential during the scattering process, we have
observed that the phase-insensitive resonant transmission is not affected.
However, the Fano-resonant backscattering, which relies on the phase
coherence in the two-channel scattering process, is completely suppressed in
this case, which is a consequence of the dephasing of individual modes due
to the interaction between them.

\section*{Acknowledgements}

We thank M. V. Fistul and V. Fleurov for useful discussions. This work was
supported by Deutsche Forschungsgemeinschaft FL200/8-1. B.A.M. appreciates
the hospitality of the Max-Planck-Institut f\"ur Physik komplexer Systeme
(Dresden) and of the Department of Physics at the Universit\"{a}t
Erlangen-N\"{u}rnberg.

\end{document}